\documentclass[journal, onecolumn, 12pt]{IEEEtran}

\usepackage{cite}
\usepackage{amsmath,amssymb,amsfonts}
\usepackage{graphicx}
\usepackage{textcomp}
\usepackage{xcolor}
\usepackage{algorithm}
\usepackage{algpseudocode}
\usepackage{float}
\usepackage{epstopdf}

\newcommand{\cmnt}[1]{\hspace*{-1px}}

\begin{document}

\title{Deep Recurrent Q-Learning Based Beam Steering Strategy for Throughput Maximization in WPCNs}
\author{
\IEEEauthorblockN{Samannaya Adhikari,~\IEEEmembership{Student Member,~IEEE}},
\and
\IEEEauthorblockN{Navchetan Awasthi~\IEEEmembership{Senior Member,~IEEE}},
\and
\IEEEauthorblockN{Siddhartha Sarma}

\thanks{S. Adhikari and S. Sarma are with the School of Computing and Electrical Engineering, Indian Institute of Technology, Mandi, Himachal Pradesh, 175005, India. N. Awasthi is with the School of Artificial Intelligence and Data Science, Indian Institute of Technology, Jodhpur, Rajasthan, 342037, Indian. Corresponding author : Samannaya Adhikari, Email: s24095@students.iitmandi.ac.in}
}

\maketitle

\begin{abstract}
In wireless powered communication networks, medium access control protocols for devices using the harvest-then-transmit strategy must be distributed, low-overhead, and capable of handling irregular and infrequent data transmissions to ensure efficient energy utilisation.
However, most existing protocols fail to meet one or more of those requirements, leading to wasted scarce harvested energy. We address this by identifying beam steering as a potential mechanism to regulate the charging rate of energy harvesting devices and thus control their access to the shared wireless medium. After formulating a joint problem of energy beam steering and slotted ALOHA-based random access, we leverage a deep learning framework based on an action-specific deep recurrent Q-Network (ADRQN) to learn a beam-steering policy only from the macro-level ternary slot outcomes, namely, idle, success and collision. Additionally, we design an oracle policy with global knowledge of the network to benchmark our proposed blind adaptive beam-steering approach. The numerical results demonstrate that our approach achieves up to 68\% increase in throughput compared to non-learning schemes, while also reaching 75-80\% of the oracle policy's performance, all without requiring channel estimation, charge-level reporting, or device-state tracking.
\end{abstract}


\IEEEpeerreviewmaketitle

\section{Introduction}\label{sec:introduction}

Recent advances and convergence in semiconductor technology, material science, and wireless communications have fueled a rapid proliferation of ultra-low power consumer electronics into our everyday lives, such as wearable devices, smart transducers, etc.
As these devices become more energy-efficient, wirelessly powering them has transitioned from a distant possibility to a near-reality. Radio Frequency energy harvesting (RFEH), a form of long-range wireless power transfer, works in a similar manner to a typical wireless communication system but facilitates the transfer of energy, instead of information, from a transmitter to a receiver~\cite{Clerckx2018}.
RFEH has the potential to transform several Internet of Things (IoT) use-cases, including healthcare, industrial automation, and environmental monitoring, smart buildings, into greener solutions by eliminating the e-waste and maintenance costs associated with batteries~\cite{Sherazi2022}. However, the integration of RFEH into conventional wireless networks gives rise to a new set of challenges involving the joint management of information and energy delivery. These must be thoroughly addressed to facilitate the widespread adoption of RFEH.

In wireless Powered Communication Networks (WPCN), a class of RFEH networks, energy harvesting (EH) devices receive energy in the downlink from\cmnt{ power beacon or} an energy transmitter and  subsequently utilise it for uplink transmission \cmnt{transmit their information in the uplink}to an access point. \cmnt{ or a base station using the harvested energy. As a result, }Unlike their battery-powered counterparts, which can transmit at regular intervals or predetermined slots, EH devices solely rely on the harvested energy for their operation,  
resulting in infrequent and irregular transmissions. Consequently, conventional Medium Access Control (MAC) protocols impose stringent constraints and operation requirements that cannot be met by EH devices. 
For instance, \emph{TDMA} requires precise coordination and frequent feedback on energy states; \emph{framed ALOHA} lacks the adaptability to varying energy availability, \emph{CSMA} and other contention-based schemes deplete scarce harvested energy on repeated channel sensing~\cite{Iannello2012}. Furthermore, advanced schemes like NOMA impose prohibitive overhead via complex encoding and precise channel estimation. These limitations necessitate the development of distributed, low-overhead MAC protocols that explicitly account for the intermittent nature of harvested energy. 

A well-designed MAC protocol must account for the unique characteristics of WPCNs to ensure the efficient utilisation of the harvested energy while maximizing the network performance. Among the class of primitive, distributed protocols, slotted ALOHA, emerges as a promising solution for meeting the unique requirements of WPCN MAC. 
It requires no channel sensing, reservation handshakes, channel estimation or synchronisation beyond slot boundaries. A device simply transmits when it has a packet ready. This simplicity aligns perfectly with the constraints of WPCNs, where communication is the primary energy-intensive task and any additional protocol overhead would drain EH device's scarce energy reserves. 

Due to the unique characteristics of the WPCN, slotted ALOHA behaves differently than in the conventional networks---uplink transmissions
are governed by the stochastic variation of wireless medium and the charging dynamics of an EH device's energy storage, rather than merely the packet arrival rate. 
This raises a natural question: \emph{can an energy transmitter regulate the transmission of uplink traffic by controlling the power delivery?} 
The answer is \emph{Yes}. Through energy beamforming, a multi-antenna energy transmitter can focus the radiated power into narrow spatial beams and steer them in different directions over time. Consequently, the charging rates of spatially distributed devices can be dynamically managed by reconfiguring the beam direction across different intervals. 

The observation reveals a vast design space for WPCN MAC protocols. 
Building on this, we pose a critical question: \emph{can a transmitter learn a beam-steering policy with the goal of maximizing the long-term network throughput through orchestrating the packet transmission solely by observing the macro-level slot outcomes, viz., a successful reception, a collision, or an idle slot?} The answer, as we demonstrate in this study, is again \emph{Yes}. 

 In this work, we propose \emph{a blind, adaptive beam steering strategy} based on an Action-specific Deep Recurrent Q-Network (ADRQN) that enables an Energy Transmitter cum Access Point (ET/AP) to learn the optimal beam steering policy without any knowledge of device locations, channel states, or charge levels. We leverage Deep Reinforcement Learning (DRL), which has shown remarkable success in learning beamforming and resource allocation policies in wireless systems under complex or partially known environments~\cite{Mismar2020, Raj2022, Wu2019, Park2022}. Our work extends these capabilities to the fundamentally different setting of WPCN, where we leverage energy beam steering as a primary mechanism for medium access control. As in our case, the agent never observes the true system state, i.e. the charge levels of the devices and channel conditions, a single slot outcome in isolation carries little information; it is the \emph{sequence} of outcomes and the actions that produced them that reveal the underlying charging patterns. For instance, a string of idle slots following a particular beam direction suggests that the devices in that direction are still charging, while a collision indicates that multiple devices have reached the transmission threshold simultaneously. The sequential dependencies necessitate a recurrent architecture that can maintain and update an internal memory across time steps.

In our preliminary work~\cite{adhikari2026adaptive}, we employed a feedforward DQN under a simplified system model, where a fully charged device could transmit regardless of the current beam direction. The present work introduces a received power threshold $P_{\mathrm{th}}$ that restricts transmission to devices within the current beam's coverage (more explained in~\ref{subsec:uplink}), tightly coupling the beam direction with transmission eligibility and providing the agent with a spatially informative feedback signal. Furthermore, we observed that the feedforward DQN, which compresses the entire history into a fixed-dimensional flattened vector, struggles to extract meaningful temporal patterns as the history length grows. This motivated the adoption of a recurrent architecture in the present work.

The key contributions of this article are as follows.

\begin{itemize}
\item We formulate the joint problem of energy beam steering and slotted ALOHA-based random access in a WPCN as a Partially Observable Markov Decision Process (POMDP), incorporating the dynamics of the charging process and its coupling with the transmission protocol.

\item We introduce a novel DRL framework in which the ET/AP learns to steer the energy beam by observing only macro-level slot outcomes---success, collision, or idle---without requiring channel estimation, charge-level feedback, or device-state tracking.

\item We design an input encoding pipeline for the ADRQN framework~\cite{zhu2017improving} tailored to the macro-level slot outcomes of the WPCN setting. Specifically, our designed architecture facilitates encoding both the previous beam index and the ternary slot outcome as one-hot vectors and projecting each through a dedicated fully connected layer with ReLU activation. We further integrate a multi-layer RNN that processes these joint embeddings to accumulate a compressed summary of the entire interaction history within its hidden state. This hidden state implicitly tracks which spatial regions have been charged, for how long, and with what outcomes, enabling the agent to make informed beam steering decisions despite lacking the global knowledge of the network. 

\item We investigate whether the macro-level feedback is a sufficient state representation for the learning agent to infer the underlying network dynamics and make effective beam steering decisions.

\item We design a model-predictive oracle policy with full state knowledge and provide a comprehensive performance analysis, benchmarking our blind approach against the oracle and non-learning baselines. We also compare the proposed recurrent architecture with a standard feedforward DQN baseline to highlight the importance of sequential history processing in this problem setting.
\end{itemize}

\subsection*{Organization and Notations}
The remainder of this paper is organized as follows:-- Section~\ref{sec:relatedworks} reviews the related literature on wireless powered communication networks, slotted ALOHA protocols, and Reinforcement Learning approaches for medium access control and resource allocation. Section~\ref{sec:system_model} presents the system model, including the beamforming, channel, energy harvesting, charging, and uplink transmission models, and formally states the problem. Section~\ref{sec:rl_solution} describes the proposed RL framework, the training procedure, and the baseline variants used for comparison. Section~\ref{sec:oracle} introduces the oracle policy with full state knowledge that serves as an upper bound for performance evaluation. Section~\ref{sec:results} presents the numerical results, including performance comparisons across different network configurations, architectural variants, and spatial device distributions. Finally, Section~\ref{sec:conclusion} concludes the paper and discusses potential directions for future work.

Throughout this paper, we adopt the following notation:-- Vectors are denoted by bold lowercase fonts (e.g., $\mathbf{w}$, $\mathbf{g}$), matrices by bold uppercase fonts (e.g., $\mathbf{I}_M$), and scalars by normal fonts. The superscripts $(\cdot)^T$, $(\cdot)^H$, and $(\cdot)^*$ denote transpose, conjugate transpose, and complex conjugate, respectively. $\mathbb{C}^{m \times n}$ and $\mathbb{R}^{m \times n}$ represent the spaces of $m \times n$ complex-valued and real-valued matrices, respectively. $\mathcal{CN}(\boldsymbol{\mu}, \boldsymbol{\Sigma})$ denotes a circularly symmetric complex Gaussian distribution with mean vector $\boldsymbol{\mu}$ and covariance matrix $\boldsymbol{\Sigma}$. $\mathbb{E}[\cdot]$ represents the expectation operator, and $\mathbb{I}\{\cdot\}$ is the indicator function.

\section{Related Works}\label{sec:relatedworks}

Early work on WPCNs adopted centralized, orthogonal (e.g., TDMA) \cite{Ju2014,Liu2014, ramezani2019optimal, liu2015throughput, lee2021residual, JuTCOM2014} or non-orthogonal (e.g., SDMA) \cite{yangJSAC2015} multi-access schemes. These foundational works proposed convex-optimisation-based designs for throughput maximization, which scale poorly with the number of devices due to their reliance on global channel information, coordinated scheduling or precise beamforming. Some protocols (e.g.~\cite{li2022demand}) require devices to explicitly request and acknowledge energy transfers, introducing additional communication overhead that consumes the very energy it seeks to harvest.



To alleviate this overhead, several works have explored distributed MAC protocols for wireless powered networks. These primarily fall into two categories: \emph{(i)} In channel sensing based MAC,~\cite{fafoutis2011odmac, nguyen2014eri} proposed transmission schemes triggered by receiver requests,~\cite{yang2012markov} proposed modified CSMA/CA protocols that incorporate device recharging dynamics, and,~\cite{Naderi2014} designed a distributed CSMA-based MAC protocol with adaptive charging thresholds for RF-powered sensors. \emph{(ii)} Parallelly, protocols based on slotted ALOHA have also been explored for their low-complexity.~\cite{ChoiShin2018} designed a slotted ALOHA-based energy harvesting MAC protocol and analytically derived the optimal number of random access slots to maximize throughput, and then a subsequent extension~\cite{Choi2019} introduced the \emph{harvest-or-access} protocol which opportunistically exploits idle slots for wireless energy transfer, improving throughput especially at high user counts,  and, recently,~\cite{gu2025exploring} applied slotted ALOHA to multi-device backscatter communications with joint beamforming and access probability optimisation. 



Several authors have explored reinforcement learning as an alternative to model-based optimisation for WPCNs.~\cite{li2018reinforcement} and~\cite{kang2020reinforcement} applied tabular Q-learning for scheduling and resource allocation in wireless powered networks. Beyond tabular methods of RL, DRL has also been applied to various problems in WPCNs. For instance,~\cite{hwang2020mult} and~\cite{sharma2019distributed} demonstrated the successful application of multi-agent DRL frameworks to distributed resource management, including time and power allocation. Close to our setting,~\cite{ahmadian2024long} proposed a multitask TD3-based DRL framework to maximise the long-term minimum ergodic throughput in a multinode WPCN, training a single centralised agent to adapt across diverse network configurations including varying node locations, battery capacities, and fading channel conditions.

In contrast to the centralised optimisation-based designs~\cite{Ju2014, Liu2014, ramezani2019optimal, liu2015throughput, lee2021residual, yangJSAC2015} that require global CSI, the participation of energy harvesting devices in channel estimation, and coordinated scheduling, and on-demand protocols~\cite{li2022demand} that consume harvested energy on explicit charging requests, our system eliminates all device-to-AP feedback. While distributed MAC protocols based on channel sensing~\cite{fafoutis2011odmac, nguyen2014eri, yang2012markov, Naderi2014} and slotted ALOHA~\cite{ChoiShin2018, Choi2019, gu2025exploring} reduce centralised coordination, they still require devices to perform non-trivial local computations such as priority-based slot selection or threshold estimation, and rely on analytical models assuming knowledge of channel statistics. On the other hand, our proposed approach requires no slot reservation, as well as all computation resides at the transmitter; the energy harvesting devices are entirely passive, simply transmitting when charged and when received power is sufficient. The RL based frameworks in~\cite{li2018reinforcement ,kang2020reinforcement, ahmadian2024long} assumes full observability with access to instantaneous battery states and channel gains of all nodes. In contrast, our ET/AP observes only a one-hot idle/success/collision signal per slot arising from slotted ALOHA contention among battery-free devices. Although in their approach,~\cite{li2018reinforcement} demonstrates that a reinforcement learning agent at the base station can learn near-optimal scheduling decisions without explicit knowledge of each node's real-time battery level, it still operates over a discretised, finite state space with collision-free TDMA access and observable queue transitions — a significantly more overhead-heavy feedback regime compared to the minimal slot-level observations used in our approach.


\section{System Model}\label{sec:system_model}
In many IoT applications, such as smart homes, smart buildings, and industrial monitoring, low-power IoT devices periodically transmit sensed data to a central coordinator. Since short-range wireless technologies like Bluetooth and WiFi typically cover tens of meters, IoT devices are placed within this range to maintain uninterrupted connectivity. Our system model follows a similar setup, where the central coordinator is replaced by an ET/AP and the IoT devices are powered through RF energy harvesting.

\begin{figure}[t]
    \centering
    \includegraphics[width=\columnwidth]{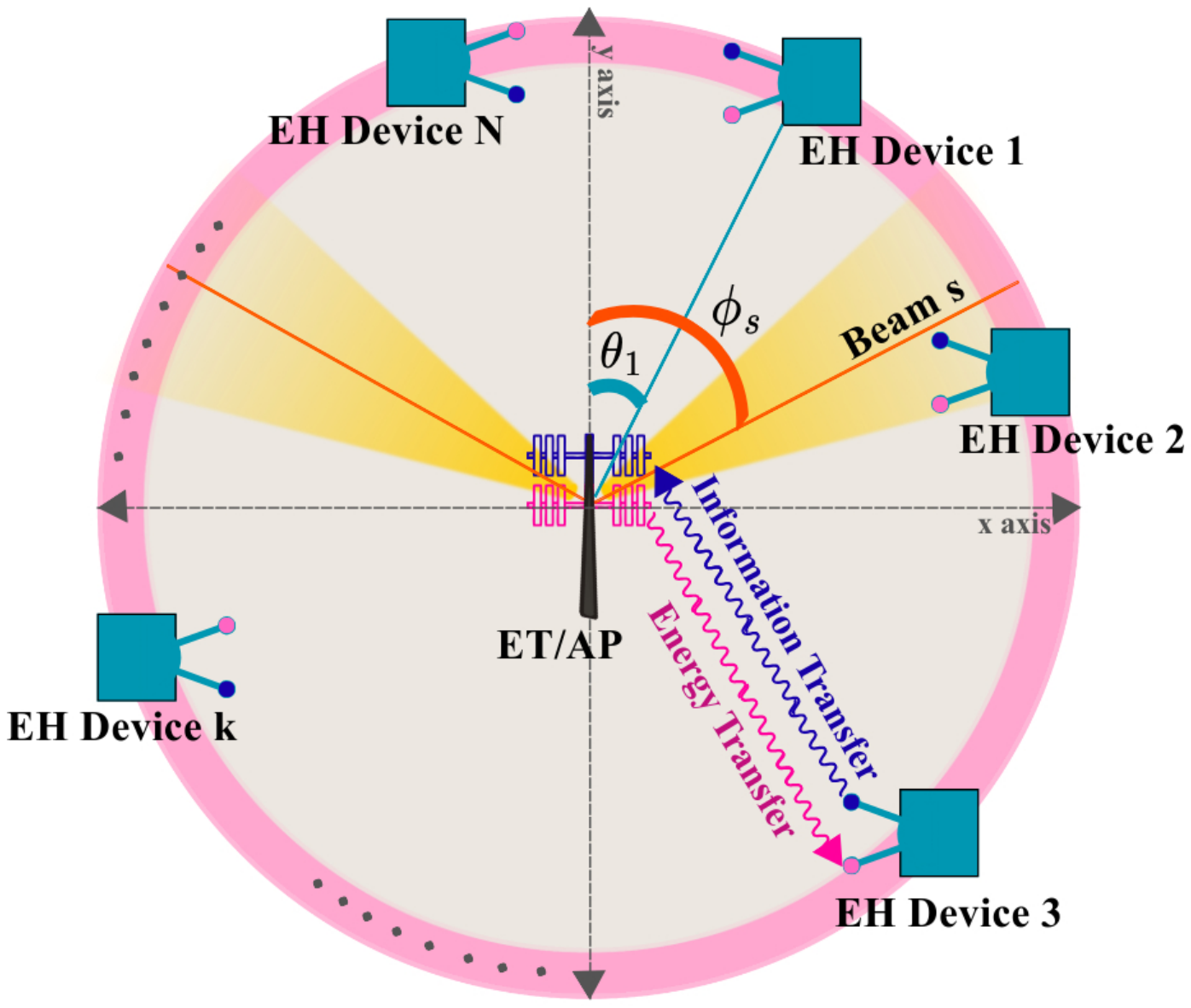}
    \caption{Pictorial depiction of the considered WPCN, where the BEHDs distributed around the ET/AP receive an energy signal in the downlink and transmit their information on the uplink once they have harvested sufficient energy.}
    \label{fig:system_model}
\end{figure}

In this work, we consider a WPCN consisting of an ET/AP with a uniform linear array (ULA) of $M$ antennas and $N$ single-antenna battery-free energy harvesting devices (BEHDs). The ET/AP performs downlink wireless energy transfer (WET) to charge the BEHDs and also receives uplink information packets via wireless information transfer (WIT) from the BEHDs in a time-slotted operation, as illustrated in Fig. \ref{fig:system_model}. The BEHDs are assumed to be identical in structure and operation and to depend solely on harvested power for their functioning. The system operates over discrete time slots with a slot duration of $\delta_t$ seconds each. We also consider that the ET/AP transmits power and information to and from the BEHDs using different frequency bands, also known as an out-of-band architecture \cite{LuX, lee2021residual}. This requires specific antennas and wireless resources at both the ET/AP and the BEHDs to transmit power and information signals, respectively. The $N$ BEHDs in this system model are assumed to be at approximately equal distances from the ET/AP; however, their exact locations are unknown to the ET/AP. This assumption is justified, particularly in single-floor or outdoor IoT sensor networks where nodes in the network tend to be at similar distances from the access point.

\subsection{Beamforming Model}\label{subsec:beamforming}

The ET/AP steers its energy beam to one of the $S$ predefined beam directions $\Phi = \{\phi_1, \phi_2, \ldots, \phi_S\}$ in each time slot and transmits with power $P_T$ to charge the BEHDs. 
The beamforming weight vector corresponding to beam direction $\phi_s$, as depicted in Fig. \ref{fig:system_model}, is given by

\begin{equation}\label{eq:beamforming}
    \mathbf{w}(\phi_s) = \frac{1}{\sqrt{M}} \left[1,\; e^{j\pi \sin(\phi_s)},\; \ldots,\; e^{j\pi (M-1) \sin(\phi_s)}\right]^T,
\end{equation}
where $\phi_s$ denotes the steering angle of the $s$-th beam.

\subsection{Channel Model}\label{subsec:channel}

The wireless channel between the ET/AP and the $i$-th BEHD, positioned at angle $\theta_i$, is modeled as a Rician fading channel. The channel vector $\mathbf{g}_i \in \mathbb{C}^{M \times 1}$ is expressed as
\begin{equation}\label{eq:channel}
    \mathbf{g}_i = \sqrt{\frac{K}{K+1}}\, \mathbf{g}_i^{\mathrm{LoS}} + \sqrt{\frac{1}{K+1}}\, \mathbf{g}_i^{\mathrm{NLoS}},
\end{equation}
where $K$ denotes the Rician $K$-factor representing the ratio of the power in the line-of-sight (LoS) component to the scattered component~\cite{xuTVT2022}. The LoS component is given by
\begin{equation}\label{eq:los}
    \mathbf{g}_i^{\mathrm{LoS}} = \sigma_l\left[1,\; e^{j\pi \sin(\theta_i)},\; \ldots,\; e^{j\pi (M-1) \sin(\theta_i)}\right]^T,
\end{equation}
where $\sigma_l^2$ denotes the path loss, and the non-line-of-sight (NLoS) component, $\mathbf{g}_i^{\mathrm{NLoS}} \sim \mathcal{CN}(\mathbf{0}, \sigma_l^2 \mathbf{I}_M)$ captures the scattered multipath contributions. The channel follows a block fading model with a coherence time $T_c$. Specifically, the channel remains constant over $\lfloor T_c / \delta_t \rfloor$ consecutive slots over a single coherence interval.

\subsection{Energy Harvesting Model}\label{subsec:harvesting}

The received power at the $i$-th BEHD in a given time slot when the ET/AP steers beam towards the direction $\phi_s$ is
\begin{equation}\label{eq:received_power}
    P_{r,i} = P_T \left| \mathbf{g}_i^H \mathbf{w}(\phi_s) \right|^2.
\end{equation}
The harvested power is calculated assuming the logistic energy harvesting function~\cite{Boshkovska}, and the harvested power at the $i$-th BEHD is modeled as
\begin{equation}\label{eq:harvest}
    P_{h,i} = \frac{\Gamma}{1 - \Omega} \left( \frac{1}{1 + e^{-\alpha(P_{r,i} - \beta)}} - \Omega \right),
\end{equation}
where $\Omega = {1}/({1 + \exp(\alpha \beta)})$, and $\Gamma$, $\alpha$, and $\beta$ are parameters characterizing the harvesting system.

\subsection{Charging Model}\label{subsec:charging}

Each BEHD stores the harvested energy in a capacitor of capacitance $C$ connected across a load resistance $R$. Since the duration of a slot is considerably less than the average time to charge the capacitor, the harvesting power varies several times over a single charging period. Therefore, a charging model, as in~\cite{sarma2025characterisation}, is adopted. In each time slot, the charge increment for device $i$ with the current charge $Q_i(t)$ and the harvested power $P_{h,i}(t)$, is given by
\begin{equation}\label{eq:charge_increment}
    \lambda_i(t) = \frac{\delta_t}{2RC} \left( -Q_i(t) + \sqrt{Q_i^2(t) + 4\, P_{h,i}(t)\, R\, C^2} \right),
\end{equation}
and the charge at the beginning of the next slot is updated as
\begin{equation}\label{eq:charge_update}
    Q_i(t+1) = \min\!\left(Q_i(t) + \lambda_i(t),\; Q_{\mathrm{th}}\right),
\end{equation}
where $Q_{\mathrm{th}}$ denotes the charge threshold at which the capacitor is considered fully charged.

\subsection{Uplink Transmission Protocol}\label{subsec:uplink}

A BEHD becomes eligible to transmit its data to the ET/AP once its stored charge reaches the threshold $Q_{\mathrm{th}}$, i.e., $Q_i(t) \geq Q_{\mathrm{th}}$. However, a fully charged BEHD transmits only if the instantaneous received power from the ET/AP exceeds a power threshold $P_{\mathrm{th}}$, i.e.,
\begin{equation}\label{eq:tx_condition}
    P_{r,i}(t) > P_{\mathrm{th}}.
\end{equation}

The threshold $P_{\mathrm{th}}$ acts as a \emph{contention-control knob}. A device whose received power exceeds $P_{\mathrm{th}}$ is admitted to contend in the slot; a device below it defers, even if fully charged. Raising $P_{\mathrm{th}}$ admits fewer devices to transmit per slot, reducing collisions, while lowering it admits more devices, reducing idle slots. Its value is just a design parameter and tuned empirically. This behaviour is confirmed numerically in Section~\ref{sec:results}, where sweeping $P_{\mathrm{th}}$ traces out the resulting trade-off between idle slots and collisions.

\begin{figure}[t]
    \centering
    \includegraphics[width=\columnwidth]{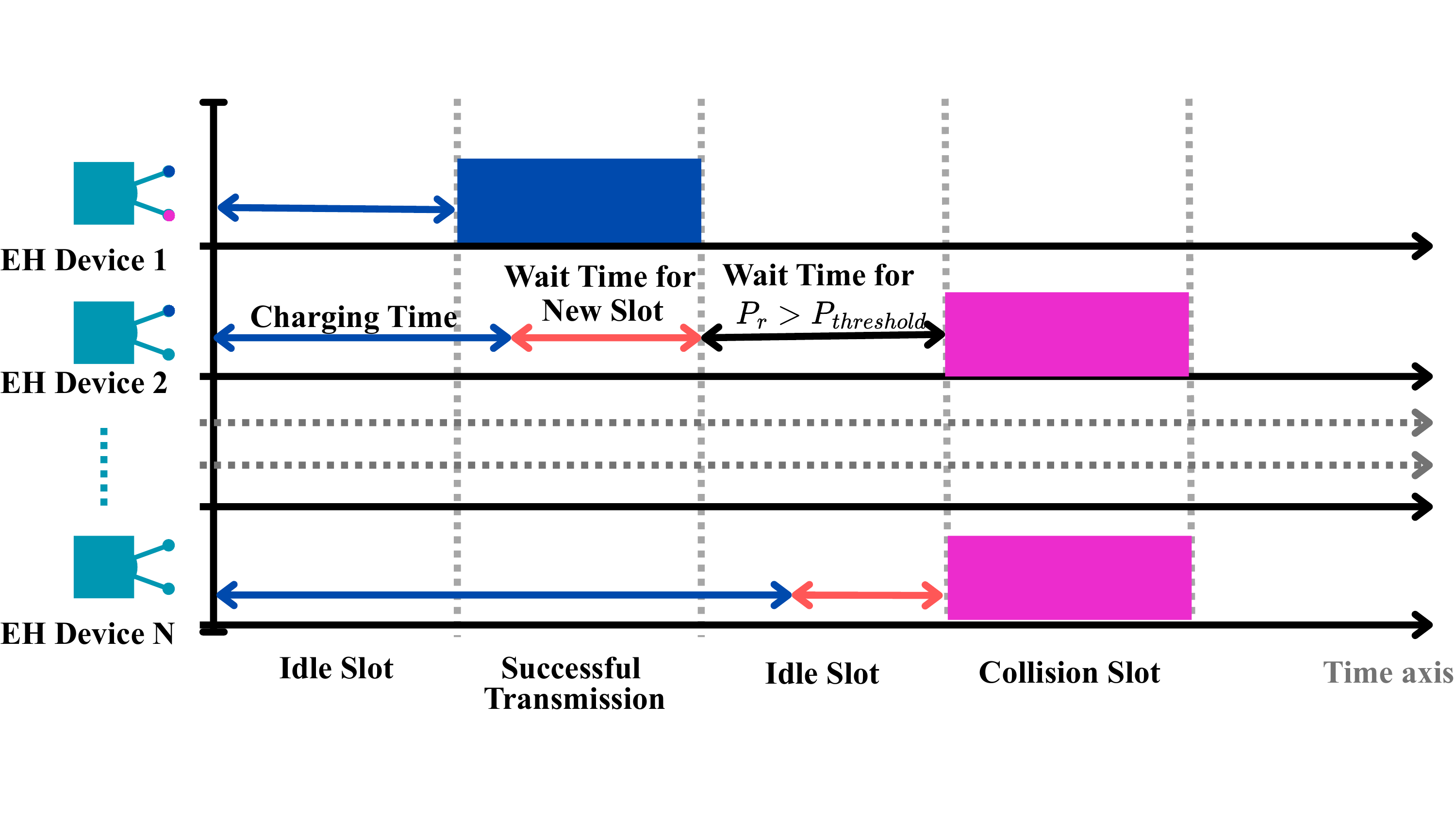}
    \caption{Slot timing diagram for the slotted ALOHA protocol, showing the charging period, the waiting time after the upper charge threshold $Q_{\mathrm{th}}$ is reached, and the three possible slot outcomes: successful transmission, idle slot, and collision.}
    \label{fig:slot_timing}
\end{figure}

The uplink channel follows a slotted ALOHA random access protocol where, in any given slot, the outcome is one of the following three possibilities:
\begin{itemize}
    \item \textbf{Idle:} No BEHD transmits, and the slot is wasted.
    \item \textbf{Success:} Exactly one BEHD transmits, and its data is successfully decoded by the ET/AP.
    \item \textbf{Collision:} Two or more BEHDs transmit simultaneously, resulting in a collision and failed reception.
\end{itemize}
The slot timing structure, depicting the charging period, the waiting time after the charge threshold is reached, and the three possible slot outcomes, is illustrated in Fig.~\ref{fig:slot_timing}.

After every transmission attempt, a BEHD's charge level is reset to the initial state $Q_0$, and the device begins a new charging cycle. 

\subsection{Problem Formulation}\label{subsec:problem_formulation}
We aim to maximize the system throughput, which is defined as the long-term average of successful packet transmissions per slot, i.e., 
\begin{equation}\label{eq:throughput}
    \eta = \lim_{T \to \infty} \frac{1}{T} \sum_{n=1}^{T} \mathbb{I}\!\left\{\text{success in slot } n\right\}.
\end{equation}


At the beginning of each slot~$n$, the ET/AP selects a beam index (action) $a(n) \in \mathcal{A} = \{1, 2, \dots, S\}$, corresponding to beam angle $\phi_{a(n)}$, and transmits energy using the steering vector $\mathbf{w}(\phi_{a(n)})$ for the entire slot duration~$\delta_t$. At the end of the slot, the ET/AP observes a feedback tuple

\begin{equation}\label{eq:obs_vector}
    \mathbf{o}(n) = \mathbf{e}_{o(n)} \in \{0,1\}^3,
\end{equation}
where $\mathbf{e}_{o(n)}$ is the one-hot encoding of the slot outcome : 
\begin{equation}\label{eq:outcome_set}
    o(n) \in \mathcal{O} = \{0\ (\text{idle}),\; 1\ (\text{success}),\; 2\ (\text{collision})\}.
\end{equation}
 The reward associated with the outcome of slot~$n$ is defined as
\begin{equation}\label{eq:reward_def}
    r(n) = \begin{cases}
    1, & \text{if } o(n) = 1,\\
    0, & \text{otherwise}.
    \end{cases}
\end{equation}
The ET/AP must learn an optimal beam selection policy $\pi\colon \mathcal{H} \to \mathcal{A}$ that maps from the observed history
\begin{equation}\label{eq:history}
\!\!\!\!\mathcal{H}(n)\! =\! \bigl\{a(0),\, \mathbf{o}(0),\, a(1),\, \mathbf{o}(1),\, \dots,\, a(n\!-\!1),\, \mathbf{o}(n\!-\!1)\bigr\}
\end{equation}
to the next beam direction index. Although the primary performance metric is the long-term average throughput~$\eta$, defined in~\eqref{eq:throughput}, for learning the optimal policy we optimise a \emph{discounted return}:
\begin{equation}\label{eq:discounted_return}
    J_\gamma(\pi) = \mathbb{E}\!\left[\sum_{n=0}^{\infty} \gamma^n\, r(n) \;\Big|\; \pi \right],
\end{equation}
where $\gamma \in [0,1)$ is the discount factor. A value of $\gamma$ close to unity guarantees that the agent will not overly favor short-term rewards and that the discounted return will act as a good surrogate objective for the long-term average throughput objective. That being said, in the numerical results \ref{sec:results} section, the final performance of the agent is measured and reported by computing the average throughput $\eta$ under the learned policy.

\section{Reinforcement Learning Based Solution}\label{sec:rl_solution}

Since the ET/AP has no access to the joint charge state $\mathbf{Q}(n)$ or the channel realisations, we therefore adopt a model-free DRL framework, in which the ET/AP acts as the learning agent. In this section, we first describe how the ADRQN framework~\cite{zhu2017improving} is tailored to the WPCN beam steering setting, detailing the architecture and training procedure, and then present two baseline variants for comparison.

%

\subsection{ADRQN-Based Beam Steering Agent}\label{subsec:adrqn}


The beam steering problem is a sequential decision process and is cast as a POMDP, defined by the tuple 
$(\mathcal{S}, \mathcal{A}, \mathcal{T}, \mathcal{O}, \Omega, r, \gamma)$, 
with the following elements.
\begin{itemize}
    \item \textbf{State space $\mathcal{S}$:} The true system state at slot~$n$ comprises the joint charge vector $\mathbf{Q}(n) \in \mathbb{R}^N$ and the instantaneous channel realisations $\{\mathbf{g}_i\}_{i=1}^{N}$. This state space is inherently high-dimensional and not directly accessible to the ET/AP, nor is a compact, sufficient representation readily available.
    \item \textbf{Action space $\mathcal{A}$:} The action is the beam index $a(n) \in \mathcal{A} = \{1, 2, \dots, S\}$, corresponding to beam direction $\phi_{a(n)} \in \Phi$.
    \item \textbf{Transition function $\mathcal{T}$:} The state transition $\mathcal{T}(s' | s, a)$ is governed by the charging dynamics~\eqref{eq:charge_increment}--\eqref{eq:charge_update}, the energy harvesting model~\eqref{eq:harvest}, and the transmission protocol in Section~\ref{subsec:uplink}.
    \item \textbf{Observation space $\mathcal{O}$:} The observation available to the agent is the slot outcome $\mathbf{o}(n) \in \{0,1\}^3$, as defined in~\eqref{eq:obs_vector}.
    \item \textbf{Observation function $\Omega$:} The mapping $\Omega(o | s', a)$ is deterministic within a coherence block: given the post transition state $s'$ and the selected beam $a(n)$, the slot outcome is uniquely determined by the number of devices that are both fully charged and satisfy~\eqref{eq:tx_condition}.
    \item \textbf{Reward:} $r(n)$ is defined in~\eqref{eq:reward_def}.
    \item \textbf{Discount factor:} $\gamma \in [0,1)$.
\end{itemize}
The objective is to maximise the discounted return $J_\gamma(\pi)$ in~\eqref{eq:discounted_return}. Since the true state is unobservable, the agent relies on the interaction history $\mathcal{H}(n)$, defined in~\eqref{eq:history}, as a surrogate for decision making.

\subsubsection{Network Architecture}\label{subsubsec:architecture}

The proposed architecture, illustrated in Fig.~\ref{fig:adrqn_arch}, consists of three modules:

\begin{figure}[t]
    \centering
    \includegraphics[width=\columnwidth]{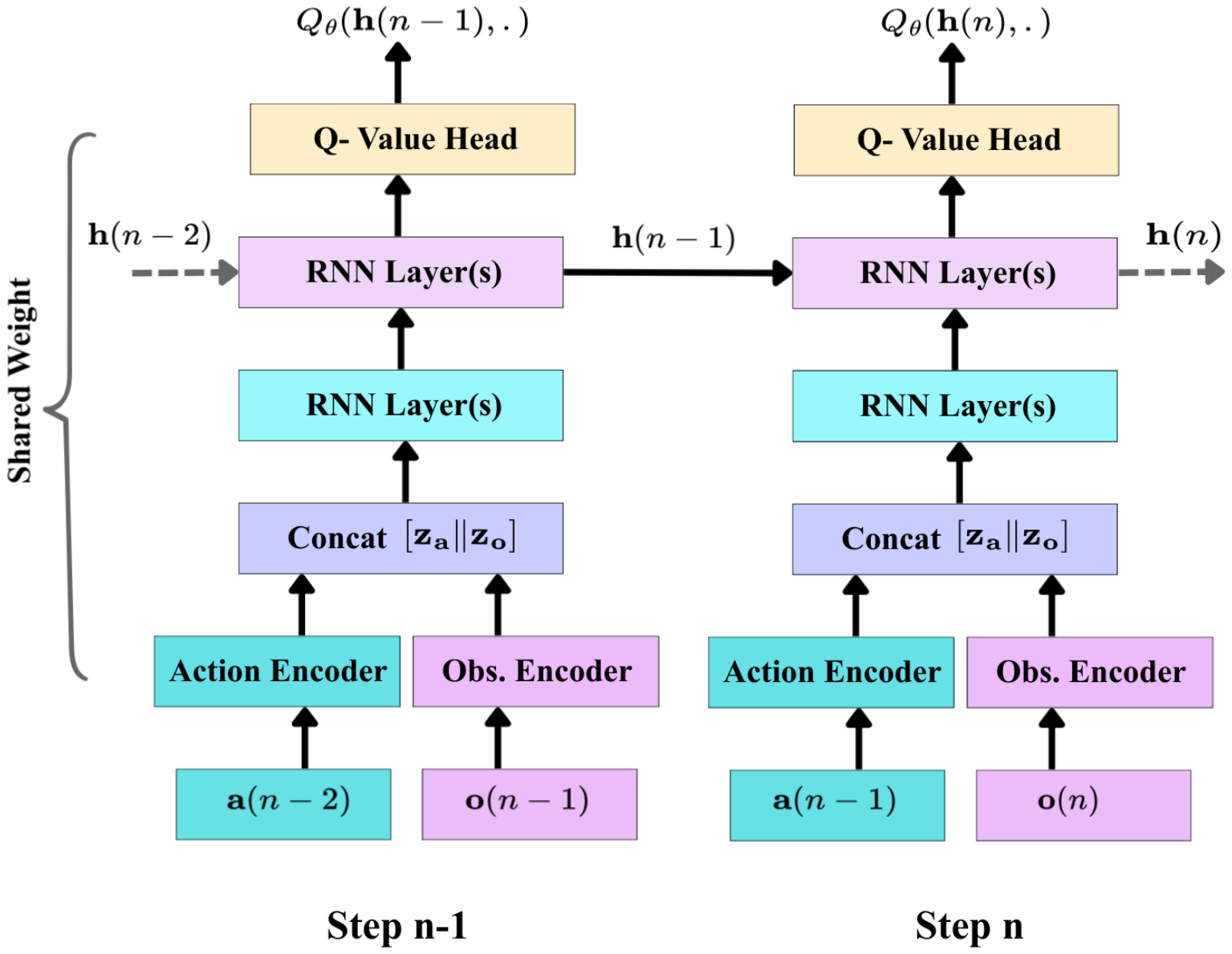}
    \caption{Architecture of the proposed ADRQN agent. The action and observation encoders produce embeddings that are concatenated and processed by a multi-layer RNN. The final hidden state is passed through a Q-value head to produce per-beam action values.}
    \label{fig:adrqn_arch}
\end{figure}

\paragraph{Action Encoder}
At each time step $n$, the previous action $a(n-1) \in \mathcal{A}$ is represented as a one-hot vector $\mathbf{a}(n-1) \in \{0,1\}^S$ and mapped to a dense embedding through a fully connected layer followed by a ReLU activation:
\begin{equation}\label{eq:action_enc}
    \mathbf{z}_a(n) = \mathrm{ReLU}\!\left(\mathbf{W}_a \,\mathbf{a}(n-1) + \mathbf{b}_a\right) \in \mathbb{R}^{d_a},
\end{equation}
where $d_a$ is the action embedding dimension.

\paragraph{Observation Encoder}
The observation $\mathbf{o}(n) \in \mathbb{R}^3$, as defined in~\eqref{eq:obs_vector}, is similarly encoded:
\begin{equation}\label{eq:obs_enc}
    \mathbf{z}_o(n) = \mathrm{ReLU}\!\left(\mathbf{W}_o \,\mathbf{o}(n) + \mathbf{b}_o\right) \in \mathbb{R}^{d_o},
\end{equation}
where $d_o$ is the observation embedding dimension.

\paragraph{Recurrent Module}
The action and observation embeddings are concatenated to form the input to a multi-layer RNN:
\begin{equation}\label{eq:rnn_input}
    \mathbf{x}(n) = \left[\mathbf{z}_a(n) \;\|\; \mathbf{z}_o(n)\right] \in \mathbb{R}^{d_a + d_o},
\end{equation}
where $\|$ denotes concatenation. The hidden state of the $l$-th RNN layer evolves as
\begin{equation}\label{eq:rnn_update}
 \!\!\!\mathbf{h}^{(l)}(n)\!= \!\tanh\!\left(\mathbf{W}_{xh}^{(l)}\,\mathbf{x}^{(l)}(n)\! +\! \mathbf{W}_{hh}^{(l)}\,\mathbf{h}^{(l)}(n-1) \!+\! \mathbf{b}_h^{(l)}\right)\!,
\end{equation}
where $\tanh(\cdot)$ is the hyperbolic tangent activation function, $\mathbf{x}^{(1)}(n) = \mathbf{x}(n)$ and $\mathbf{x}^{(l)}(n) = \mathbf{h}^{(l-1)}(n)$ for $l \geq 2$, and $L$ denotes the total number of recurrent layers. The hidden state $\mathbf{h}^{(L)}(n) \in \mathbb{R}^{d_h}$ of the final layer encodes a compressed representation of the entire interaction history up to slot~$n$.

\paragraph{Q-Value Head}
The Q-values for all beam directions are computed from the top-layer hidden state via a two-layer feedforward network:
\begin{equation}\label{eq:q_head}
    Q_\theta\!\left(\mathcal{H}(n), a\right) = \left[\mathbf{W}_2 \,\mathrm{ReLU}\!\left(\mathbf{W}_1 \,\mathbf{h}^{(L)}(n) + \mathbf{b}_1\right) + \mathbf{b}_2\right]_a,
\end{equation}
where the subscript $a$ selects the component corresponding to action $a \in \mathcal{A}$, and $\theta$ collectively denotes all trainable parameters of the network.

\subsubsection{Action Selection}
During interaction with the environment, the agent follows an $\varepsilon$-greedy policy. At slot~$n$, the action is selected as
\begin{equation}\label{eq:epsilon_greedy}
    a(n) = \begin{cases}
    \text{Uniform}(\mathcal{A}), & \text{with probability } \varepsilon, \\
    \arg\max_{a \in \mathcal{A}}\, Q_\theta\!\left(\mathcal{H}(n), a\right), & \text{otherwise},
    \end{cases}
\end{equation}
where the exploration rate $\varepsilon$ is decayed multiplicatively after each episode as $\varepsilon \leftarrow \max(\varepsilon_{\min},\; \varepsilon \cdot \varepsilon_{\mathrm{decay}})$.

\subsection{Training Procedure}\label{subsec:training}

The ADRQN stores complete episodes in a sequence replay buffer. During training, a minibatch of $B$ sequences, each of length $H$, is sampled by selecting a random episode and extracting a contiguous subsequence from a uniformly random starting position. To address the mismatch between the zero-initialised hidden state at the start of a sampled subsequence and the hidden state that would have resulted from the full preceding history, we adopt a \emph{burn-in} strategy~\cite{kapturowski2018recurrent}: each sequence is partitioned into a burn-in prefix of length $H_b$ and a training suffix of length $H - H_b$, where the prefix is processed without computing gradients solely to produce a warm hidden state. Training updates are performed every $U$ environment steps. The TD loss is computed over the training suffix only. At training step~$n$ within the suffix, the policy network processes $(\mathbf{a}(n-1), \mathbf{o}(n))$ to produce Q-values $Q_\theta(\mathbf{h}(t), \cdot)$, and the TD target is
\begin{equation}\label{eq:td_target}
    y(n) = r(n) + \gamma \max_{a' \in \mathcal{A}} Q_{\bar{\theta}}\!\left(\bar{\mathbf{h}}(n\!+\!1), a'\right)\left(1 - d(n)\right),
\end{equation}
where $\bar{\mathbf{h}}(n+1)$ is obtained by feeding $(\mathbf{a}(n), \mathbf{o}(n+1))$ into the target network. 
Here, $d(n) \in \{0, 1\}$ is the terminal flag, which equals~$1$ if slot~$n$ is the last slot of the episode and~$0$ otherwise. The loss over the training suffix is
\begin{equation}\label{eq:loss}
    \mathcal{L}(\theta) = \frac{1}{H - H_b} \sum_{n=H_b}^{H-1} \left(y(n) - Q_\theta\!\left(\mathbf{h}(n), a(n)\right)\right)^2.
\end{equation}
The target network parameters $\bar{\theta}$ after every training step are updated via :
\begin{equation}\label{eq:soft_update}
    \bar{\theta} \leftarrow \tau\, \theta + (1 - \tau)\, \bar{\theta},
\end{equation}
with gradient clipping applied to prevent exploding gradients. The complete procedure is summarised in Algorithm~\ref{alg:adrqn}.

\begin{algorithm}[h]
\caption{ADRQN-Based Beam Steering Agent for WPCN}
\label{alg:adrqn}
\begin{algorithmic}[1]
\State Initialise policy network $Q_\Psi$ and target network $Q_{\bar{\Psi}}$ with $\bar{\Psi} \leftarrow \Psi$
\State Initialise sequence replay buffer $\mathcal{D}$, exploration rate $\varepsilon \leftarrow 1.0$
\For{episode $= 1, 2, \dots, E$}
    \State Initialise hidden state $\mathbf{h} \leftarrow \mathbf{0}$, episode buffer $\mathcal{T} \leftarrow \emptyset$
    \For{slot $n = 0, 1, \dots, T-1$}
        \State Select action $a(n)$ via $\varepsilon$-greedy policy~\eqref{eq:epsilon_greedy} 
        \State Apply $a(n)$ to the environment; 
        \State observe outcome $\mathbf{o}(n)$ and reward $r(n)$
        \State Append $(\mathbf{a}(n-1), \mathbf{o}(n-1), a(n), r(n), \mathbf{o}(n), d(n))$ to $\mathcal{T}$
        \If{$n \bmod U = 0$ \textbf{and} $|\mathcal{D}| \geq B$}
            \State Sample minibatch of $B$ sequences of length $H$ from $\mathcal{D}$
            \State Burn-in: forward the first $H_b$ steps without gradient
            \State Compute loss $\mathcal{L}(\Psi)$ over remaining $H - H_b$ steps via~\eqref{eq:loss}
            \State Update $\Psi$ via Adam with gradient clipping
            \State Soft-update target: $\bar{\Psi} \leftarrow \tau\,\Psi + (1 - \tau)\,\bar{\Psi}$
        \EndIf
    \EndFor
    \State Store episode $\mathcal{T}$ in replay buffer $\mathcal{D}$
    \State Decay $\varepsilon \leftarrow \max(\varepsilon_{\min},\; \varepsilon \cdot \varepsilon_{\mathrm{decay}})$
\EndFor
\end{algorithmic}
\end{algorithm}

\subsection{Baseline Variants}\label{subsec:baselines}

To highlight the contribution of the recurrent architecture and to study the effect of different temporal modelling choices, we consider two baseline variants that share the same training hyperparameters as the proposed ADRQN, differing only in the mechanism used to process temporal information.

\subsubsection{ADRQN-LSTM}\label{subsubsec:lstm}

The first variant replaces the Elman RNN in the recurrent module with a Long Short-Term Memory (LSTM) network~\cite{hochreiter1997long}, which employs forget, input, and output gates to selectively retain or discard information over long sequences. This gating mechanism may be beneficial when the actual dynamics of the system span many time slots. The LSTM can better propagate relevant information such as the approximate charging phase of device clusters over extended horizons. The LSTM maintains a hidden state tuple $(\mathbf{h}(n), \mathbf{c}(n))$, where $\mathbf{h}(n)$ is the output hidden state and $\mathbf{c}(n)$ is the cell state which is an internal memory vector that is selectively updated by the gating mechanisms at each step, enabling the network to retain or forget information over long time horizons. Both components are warmed up during the burn-in phase. All other aspects of the architecture and training procedure remain identical to the RNN-based ADRQN described in Algorithm~\ref{alg:adrqn}.

\subsubsection{Feedforward DQN with History Window}\label{subsubsec:ff_dqn}
The second baseline uses a standard Deep Q-Network (DQN)~\cite{mnih2015human}, in which a feedforward neural network $Q_\theta$ approximates the optimal action-value function. The parameters are updated by minimising the temporal-difference loss
\begin{equation}\label{eq:td_loss}
    \mathcal{L}(\theta) = \mathbb{E}\!\left[\left(r + \gamma \max_{a'} Q_{\bar{\theta}}(s', a') - Q_\theta(s, a)\right)^2\right],
\end{equation}
where $\bar{\theta}$ denotes the target network parameters. Since the DQN requires a fixed-dimensional state input, partial observability is addressed by concatenating the most recent $H$ action--observation pairs into a single vector:
\begin{equation}\label{eq:dqn_state}
    \mathbf{s}(n)\! =\! \left[\mathbf{a}(n\!-\!H) \| \mathbf{o}(n\!-\!H) \|\! \cdots\! \| \mathbf{a}(n\!-\!1) \| \mathbf{o}(n\!-\!1)\right]\! \in\! \mathbb{R}^{H(S + 4)}.
\end{equation}
This vector is processed by a feedforward encoder comprising two fully connected layers with ReLU activations:
\begin{equation}\label{eq:dqn_encoder}
    \mathbf{z}(n) = \mathrm{ReLU}\!\left(\mathbf{W}_2'\, \mathrm{ReLU}\!\left(\mathbf{W}_1'\, \mathbf{s}(n) + \mathbf{b}_1'\right) + \mathbf{b}_2'\right),
\end{equation}
followed by the same Q-value head as in~\eqref{eq:q_head}. A standard experience replay buffer storing individual transitions $(\mathbf{s}(n), a(n), r(n), \mathbf{s}(n+1), d(n))$ is used.

\section{Oracle Upper Bound}\label{sec:oracle}

 To establish a performance ceiling against which the proposed learning-based policies can be benchmarked, we design an \emph{oracle} beam steering policy that assumes the ET/AP has perfect knowledge of the instantaneous charge levels $\mathbf{Q}(n)$, the current channel realisations $\{\mathbf{g}_i\}_{i=1}^{N}$, the beam coverage geometry, and the thresholds $Q_{\mathrm{th}}$ and $P_{\mathrm{th}}$. Since the slot outcome under any candidate beam can be computed deterministically from this information within a coherence block, the oracle does not learn through exploration; it directly evaluates the consequences of its actions via forward simulation and selects the best one.

For a given beam index $a \in \mathcal{A}$, in the slot~$n$, the \emph{coverage set} contains all devices whose received power exceeds the threshold:
\begin{equation}\label{eq:coverage_set}
    \mathcal{C}_a(n) = \left\{i \in \{1, \dots, N\} : P_T \left|\mathbf{g}_i^H \mathbf{w}(\phi_a)\right|^2 > P_{\mathrm{th}}\right\}.
\end{equation}
A device is said to be \emph{covered} by beam $\phi_s$ if it belongs to $\mathcal{C}_a(n)$, i.e., its received power under that beam exceeds $P_{\mathrm{th}}$. The \emph{ready set} is the subset of covered devices that are also fully charged:
\begin{equation}\label{eq:ready_set}
    \mathcal{R}_a(n) = \left\{i \in \mathcal{C}_a(n) : Q_i(n) \geq Q_{\mathrm{th}}\right\}.
\end{equation}
The cardinality $|\mathcal{R}_s(n)|$ directly determines the slot outcome: idle if $|\mathcal{R}_a(n)| = 0$, success if $|\mathcal{R}_a(n)| = 1$, and collision if $|\mathcal{R}_a(n)| \geq 2$. Within a coherence block, the coverage sets remain constant 

\subsection{Policy}\label{subsec:oracle_policy}

The oracle operates in two tiers, evaluated sequentially at each slot.

\subsubsection{Tier~1: Immediate Success}

The oracle first checks whether any beam yields an immediate success. Let $\mathcal{S}_1(n) = \{a \in \mathcal{A} : |\mathcal{R}_a(n)| = 1\}$. If $|\mathcal{S}_1(n)| \geq 1$, the oracle selects a beam from this set, bypassing the more expensive sequence search. When multiple beams qualify, ties are broken by simulating the charge update and device reset for each candidate $a \in \mathcal{S}_1(n)$ and computing the worst-case collision potential at the next slot as $\rho(a) = \max_{a' \in \mathcal{A}} |\hat{\mathcal{R}}_{a'}^{(a)}(n+1)|$, where $\hat{\mathcal{R}}_{a'}^{(a)}(n+1)$ is the prospective ready set after selecting beam $a$ and resetting the transmitting device. The beam with smallest $\rho(a)$ is chosen.

\subsubsection{Tier~2: Exhaustive Sequence Search}

When $\mathcal{S}_1(n) = \emptyset$, the oracle searches over all $S^k$ beam sequences of length $k$. For each candidate sequence $\mathbf{a} = (a_0, a_1, \dots, a_{k-1}) \in \mathcal{A}^k$, the system dynamics are simulated starting from the current state $\mathbf{Q}(n)$. At each simulated step $t \in \{0, \dots, k-1\}$, the charges are updated via~\eqref{eq:charge_increment}--\eqref{eq:charge_update} under beam $a_t$, the ready set $\mathcal{R}_{a_t}(n+t)$ is computed, and the outcome is recorded as
\begin{equation}\label{eq:sigma}
    \sigma_t(\mathbf{a}) = \mathbb{I}\!\left\{|\mathcal{R}_{a_t}(n+t)| = 1\right\}.
\end{equation}
If $|\mathcal{R}_{a_t}(n+t)| \geq 1$ (success or collision), the charges of all transmitting devices are reset to $Q_0$. The total number of successes over the horizon is
\begin{equation}\label{eq:total_successes}
    \Sigma(\mathbf{a}) = \sum_{t=0}^{k-1} \sigma_t(\mathbf{a}).
\end{equation}
The oracle selects the first action of the sequence that maximises $\Sigma(\mathbf{a})$:
\begin{equation}\label{eq:mpc_select}
    a^*(n) = \text{first}\!\left(\arg\max_{\mathbf{a} \in \mathcal{A}^k} \Sigma(\mathbf{a})\right).
\end{equation}
Ties among sequences with the same $\Sigma^*$ are broken hierarchically: first, the sequence whose earliest success occurs at the smallest step index is preferred; if still tied, the sequence whose terminal charge state $\hat{\mathbf{Q}}^{(\mathbf{a})}(n+k)$ contains the fewest devices with charge exceeding $\xi \cdot Q_{\mathrm{th}}$ is selected, where $\xi \in (0,1)$ is a near-threshold parameter. This favours actions that leave the system in the least collision-prone state beyond the planning horizon.

The complete oracle policy is thus
\begin{equation}\label{eq:oracle_policy}
    a^*(n) = \begin{cases}
    \displaystyle\arg\min_{s \in \mathcal{S}_1(n)} \rho(s), & \text{if } \mathcal{S}_1(n) \neq \emptyset, \\[8pt]
    \text{first}\!\left(\displaystyle\arg\max_{\mathbf{a} \in \mathcal{A}^k} \Sigma(\mathbf{a})\right), & \text{otherwise},
    \end{cases}
\end{equation}
and is summarised in Algorithm~\ref{alg:oracle}.

\begin{algorithm}[t]
\caption{Oracle Beam Steering Policy (MPC)}\label{alg:oracle}
\begin{algorithmic}[1]
\Require State $\mathbf{Q}(n)$, channels $\{\mathbf{g}_i\}$, beams $\{\mathbf{w}(\phi_a)\}_{a=1}^{S}$, thresholds $Q_{\mathrm{th}}$, $P_{\mathrm{th}}$, horizon $k$, parameter $\xi$
\For{each slot $n$}
    \State Compute $P_r(a,i)$, $\mathcal{C}_a(n)$ for all $a$
    \State Compute $\mathcal{R}_a(n)$ for all $a \in \mathcal{A}$
    \State $\mathcal{S}_1(n) \leftarrow \{a : |\mathcal{R}_a(n)| = 1\}$
    \If{$\mathcal{S}_1(n) \neq \emptyset$} \hfill \textit{(Tier 1: immediate success)}
        \For{each $a \in \mathcal{S}_1(n)$}
            \State Simulate charge update and reset; 
            \State compute $\rho(a)$
        \EndFor
        \State Select $a^*(n) = \arg\min_{a \in \mathcal{S}_1(n)} \rho(a)$
    \Else \hfill \textit{(Tier 2: exhaustive sequence search)}
        \For{each sequence $\mathbf{a} \in \mathcal{A}^k$}
            \State $\hat{\mathbf{Q}} \leftarrow \mathbf{Q}(n)$, $\Sigma \leftarrow 0$
            \For{$t = 0, 1, \dots, k-1$}
                \State Update $\hat{\mathbf{Q}}$ via~\eqref{eq:charge_increment}--\eqref{eq:charge_update} under beam $a_t$
                \State Compute $\mathcal{R}_{a_t}$; if $|\mathcal{R}_{a_t}| = 1$ then $\Sigma \leftarrow \Sigma + 1$
                \State If $|\mathcal{R}_{a_t}| \geq 1$: reset $\hat{Q}_i \leftarrow Q_0$ for $i \in \mathcal{R}_{a_t}$
            \EndFor
            \State Record $(\Sigma, \text{first success step}, \hat{\mathbf{Q}})$ for $\mathbf{a}$
        \EndFor
        \State Select $a^*(n) = $ first action of best $\mathbf{a}$ via~\eqref{eq:mpc_select} with  \hspace*{3em}tie-breaking
    \EndIf
    \State Execute beam $a^*(n)$; 
    \State update $\mathbf{Q}$ and reset transmitting devices
\EndFor
\end{algorithmic}
\end{algorithm}
\section{Numerical Results}\label{sec:results}

\subsection{Simulation Settings}\label{subsec:setup}

All simulations are performed over $E = 900$ episodes, where each episode is divided into $T = 3500$ time slots. we perform a few independent experiments, each with a different random device placement where the angular position of each BEHD is drawn independently and uniformly from $[0^\circ, 360^\circ)$. The device placements are kept identical across all algorithms within each experiment to ensure a fair comparison. The reported \emph{average throughput} values are calculated by first averaging the throughput over the last 100 episodes of each simulation run and then averaging over all independent device placement configurations. The architectures for the recurrent models consist of embedding dimensions for actions and observations, which are set to 32. For the oracle policy, the planning horizon is set to $k = 5$ and the near-threshold parameter to $\xi = 0.85$. All architectures share a hidden dimension of 128 and the \emph{Adam optimizer} is utilized for training keeping the learning rate at $10^{-5}$. The target networks are updated via Polyak averaging with coefficient $\tau = 0.005$, and gradient norms are clipped to a maximum of 10. The exploration rate is initialised to $\varepsilon = 1.0$ and decayed multiplicatively by a factor of 0.995 per episode to a minimum of 0.05. For the recurrent architectures, the sequence replay buffer stores entire episodes, and the capacity is 10{,}000. A burn-in length of $H_b = 10$ is used for hidden state warm-up. Training updates are performed every 6 environment steps with a minibatch size of 32.


\subsection{System Parameters}\label{subsec:params}

The physical layer parameters used in all simulations are summarised in Table~\ref{tab:params}. 
\begin{table}[t]
\centering
\caption{System Parameters}
\label{tab:params}
\renewcommand{\arraystretch}{1.15}
\begin{tabular}{ll|ll|ll}
\hline
\textbf{Param.} & \textbf{Value} & \textbf{Param.} & \textbf{Value} & \textbf{Param.} & \textbf{Value} \\
\hline
$K$          & 6 dB          & $P_T$        & 1 W            & $\delta_t$     & 0.1 s        \\
$\Gamma$     & 0.024 W       & $C$          & 1 mF           & $T_c$          & 1.0 s        \\
$\alpha$     & 150           & $R$          & 100 $\Omega$   & $\sigma_l^2$   & $10^{-2}$    \\
$\beta$      & 0.014         & $Q_0$        & 1.5 mC         & $P_{\mathrm{th}}$ & $-12$ dB  \\
$\gamma$     & 0.95          & $Q_{\mathrm{th}}$ & 3.0 mC   & $T$            & 3500 slots   \\
\hline
\end{tabular}
\end{table}
The set of steering angles $\{\phi_s\}_{s=1}^{S}$ is chosen such that, for a given number of antennas $M$ and power threshold $P_{\mathrm{th}}$, the union of the beam coverage regions spans nearly the full $360^\circ$ angular range with minimal overlap between adjacent lobes. This design criterion serves two purposes: it ensures that every BEHD, regardless of its angular position, falls within the coverage of at least one beam, and the limited overlap reduces the likelihood of multiple beams simultaneously charging the same set of devices, thereby helping the agent learn effective spatial scheduling patterns. 

\begin{figure}[t]
    \centering
    \includegraphics[width=\columnwidth]{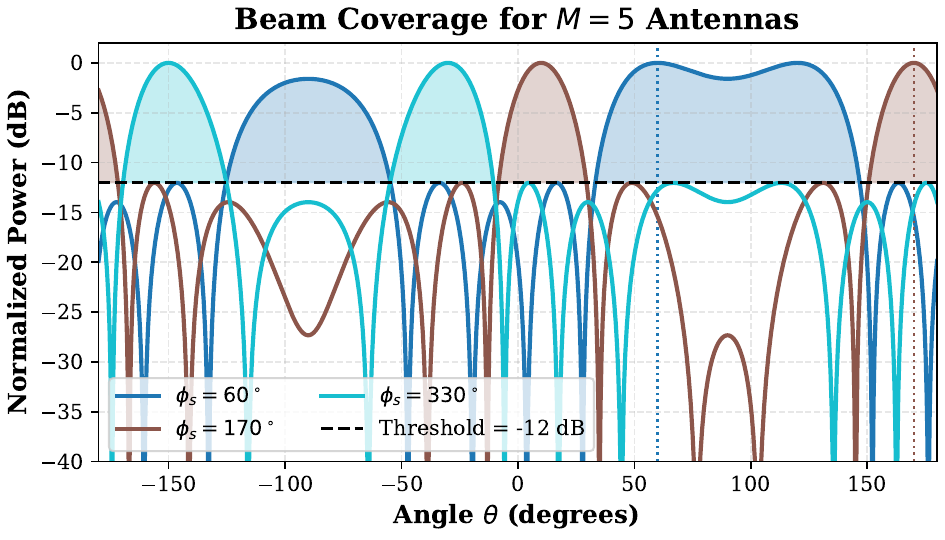}
    \caption{Beam coverage for $M = 5$ antennas with $S = 3$ steering directions $\{\phi_s\} = \{60^\circ, 170^\circ, 330^\circ\}$. The shaded regions indicate angular ranges where the normalised received power exceeds the threshold $P_{\mathrm{th}} = -12$~dB.}
    \label{fig:coverage_5}
\end{figure}

\begin{figure}[t]
    \centering
    \includegraphics[width=\columnwidth]{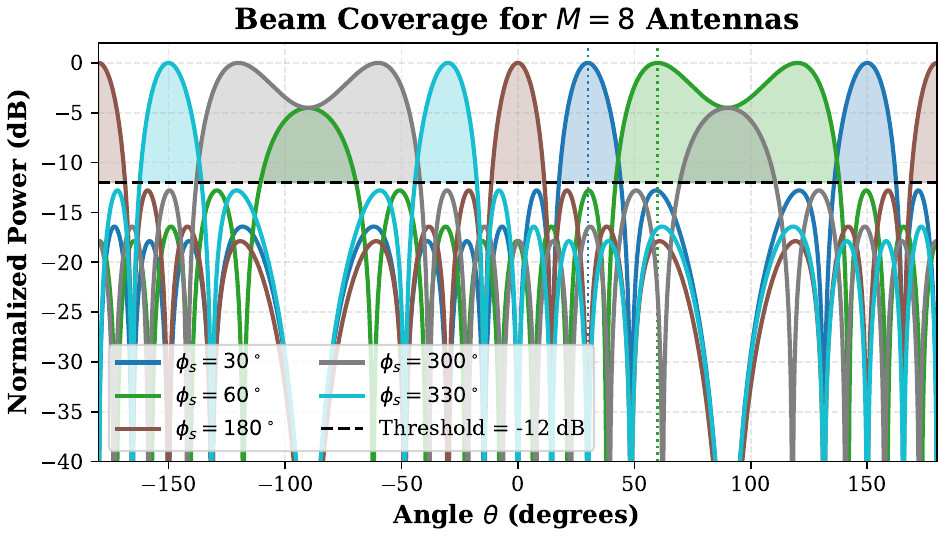}
    \caption{Beam coverage for $M = 8$ antennas with $S = 5$ steering directions $\{\phi_s\} = \{30^\circ, 60^\circ, 180^\circ, 300^\circ, 330^\circ\}$. The shaded regions indicate angular ranges where the normalised received power exceeds the threshold $P_{\mathrm{th}} = -12$~dB.}
    \label{fig:coverage_8}
\end{figure}

To determine the beam directions, we compute the normalised array factor for each candidate steering angle and identify the angular regions where the received power exceeds $P_{\mathrm{th}}$. We repeat the same process for different sets of steering angles and find the set which approximately covers the whole $360^{\circ}$ with minimal overlap among the lobes. Figs.~\ref{fig:coverage_5} and~\ref{fig:coverage_8} illustrate the beam coverages with  two of such beam directions sets for $M = 5$ and $M = 8$ antennas, respectively.
For the 5-antenna configuration, three steering directions $\Phi\ = \{60^\circ, 170^\circ, 330^\circ\}$ are taken and for the 8-antenna configuration, due to the narrower beamwidth, five directions $\{\phi_s\} = \{30^\circ, 60^\circ, 180^\circ, 300^\circ, 330^\circ\}$ are considered. In the remainder of this section, unless stated otherwise, the 5-antenna experiments use $S = 3$ beams and the 8-antenna experiments use $S = 5$ beams as defined above.

\subsection{Performance Analysis}\label{subsec:perf}

\begin{figure}[t]
    \centering
    \includegraphics[width=\columnwidth]{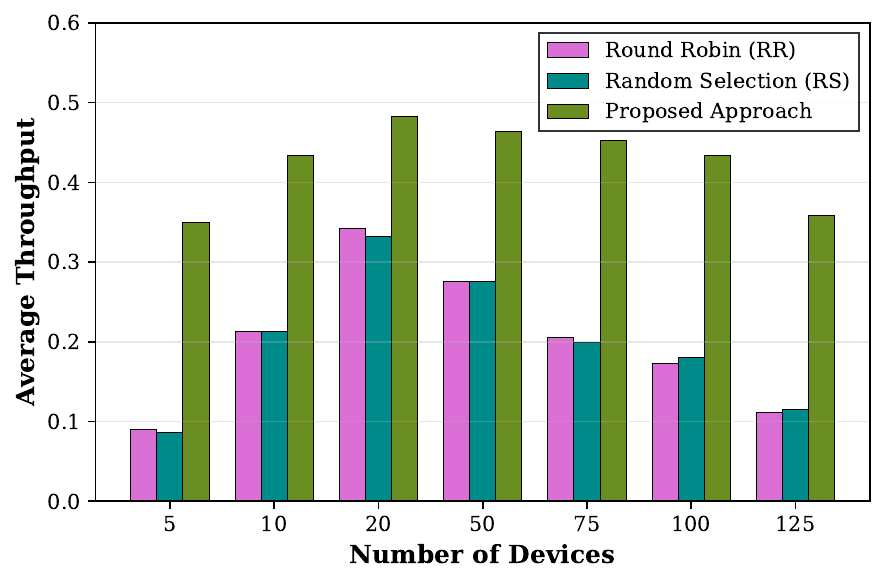}
    \caption{Average throughput (averaged over all device placement configurations for the last 100 episodes) versus number of BEHDs for the proposed ADRQN ($H = 50$, $M = 8$), Round Robin (RR), and Random Selection (RS).}
    \label{fig:baseline_comparison}
\end{figure}

Fig.~\ref{fig:baseline_comparison}  shows the comparison between the proposed ADRQN-based beam steering policy (with history length $H = 50$ and $M = 8$ antennas) and two non-learning baselines, namely Round Robin (RR) and Random Selection (RS). The performance of the two baselines is evaluated over the network size ranging from  $N = 5$ to $N = 125$ BEHDs. In the RR policy, the ET/AP sequentially selects $S$ predetermined steering vectors and chooses one per slot in a fixed order. In the RS policy, the beam direction is selected uniformly at random per slot. Both the RR and RS policies do not utilize the charge state and spatial location information of the devices; hence, they provide fair baselines to compare the performance of the proposed learning-based policy. The performance of the RR and RS obtained through simulating the same device placement configurations as the evaluation of the proposed policy.

The results show that the proposed policy outperforms the two baselines in all network configurations. For example, at $N = 50$ devices, the proposed approach achieves an average throughput of $0.4635$, whereas the Round Robin and Random Selection policies achieve throughputs of $0.2758$ and $0.2755$, respectively. Hence, the proposed policy achieves around $68\%$ improvement compared to the two baselines. As the network size increases from $N = 5$ to a moderate value, the throughput of all three algorithms increases. This is because more the number of devices enable the effective utilization of the idle time slots. However, beyond a certain network density, the throughput decreases due to the increased number of collisions among the fully charged devices using the slotted ALOHA protocol.

\begin{figure}[t]
    \centering
    \includegraphics[width=\columnwidth]{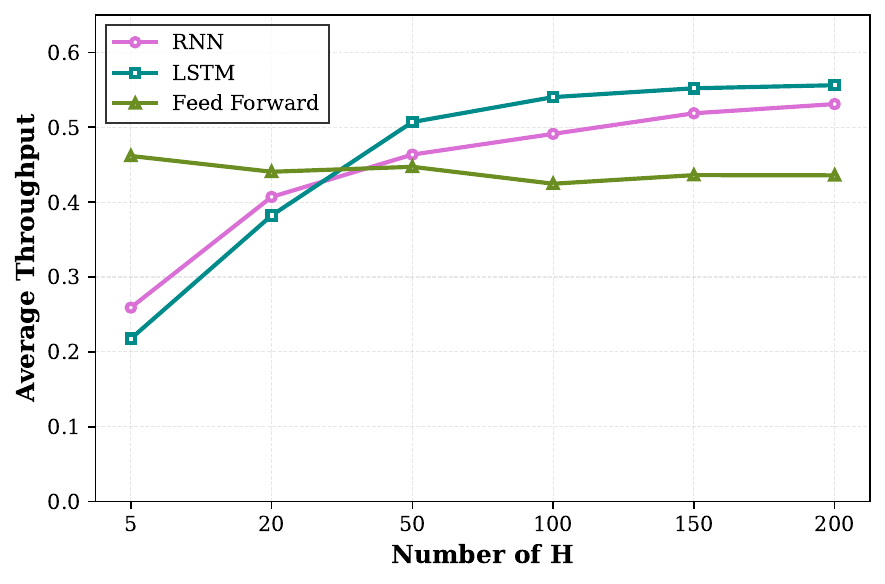}
    \caption{Average throughput (averaged over all device placement configurations for the last 100 episodes) versus history/sequence length $H$ for the feedforward DQN, ADRQN (RNN), and ADRQN (LSTM) with $N = 50$ and $M = 8$.}
    \label{fig:arch_comparison}
\end{figure}

A comparison of the three different neural network architectures' throughput as a function of the sequence length $H$ is depicted in Fig.~\ref{fig:arch_comparison}, which considers a neural network consisting of  $N = 50$ BEHDs and $M = 8$ antennas. The feedforward DQN achieves competitive results when the sequence length is relatively short (e.g., $H = 10$ ), but the throughput gets reduced as $H$ is increased. This is because the input dimensionality of the feedforward DQN grows linearly with $H$, given by $H(S + 4)$, and when $H$ is large, the input dimensionality is too large for the fixed hidden layer dimensionality, making optimisation increasingly difficult.  Both the recurrent architectures  exhibit increasing throughput with increasing $H$, and this is because the fixed hidden dimensionality of the RNN and LSTM is not affected by the sequence length $H$, allowing them to effectively capture patterns in the sequence of received signals over long sequence lengths. The LSTM performs slightly better than the RNN over long sequence lengths, which is expected given the superior capacity of the LSTM to retain information over long time scales. 

\begin{figure}[t]
    \centering
    \includegraphics[width=\columnwidth]{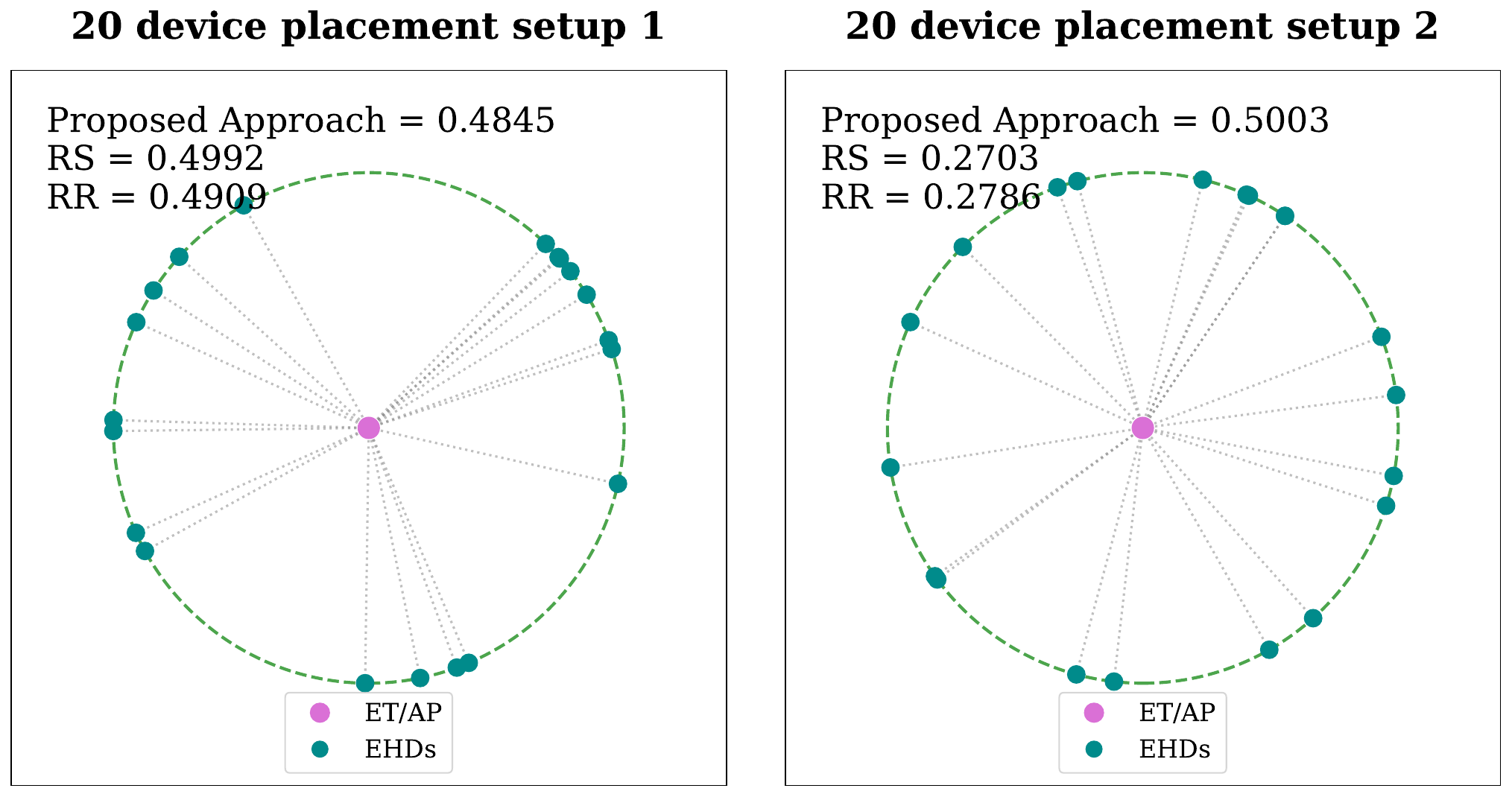}
    \caption{Two representative device placement configurations for $N = 20$ BEHDs. Setup~1 (left): RR and RS achieve throughput comparable to the ADRQN. Setup~2 (right): RR and RS degrade significantly while the ADRQN maintains robust performance.}
    \label{fig:placement_20}
\end{figure}

Figs.~\ref{fig:placement_20} shows the sensitivity of beam steering algorithms in terms of their response to the spatial distribution of BEHDs in networks with $N = 20$ devices. Two representative configurations are shown from among the different independent placements, with one configuration where the performance (i.e. the throughput averaged over the last 100 episodes out of 900 episodes for the shown device placement setup only) of RR and RS is comparable to that of the proposed approach, and another where their performance is significantly lower than the proposed approach. For the $20$ device placement scenario, in setup 1 (left panel of Fig.~\ref{fig:placement_20}), we have throughput  0.4909 for RR and 0.4992 for RS, which is comparable to the throughput of 0.4845 for ADRQN based algorithm. However, in setup~2 (right panel), the performance of both RR and RS drops significantly to 0.2787 and 0.2703, respectively, while ADRQN  based algorithm achieves a throughput of 0.5003. This shows that the proposed approach achieves high performance regardless of the spatial distribution of BEHDs, while both non-adaptive approaches exhibit significant sensitivity in response to the relative positioning of devices with respect to fixed beam directions.


\begin{figure}[t]
    \centering
    \includegraphics[width=\columnwidth]{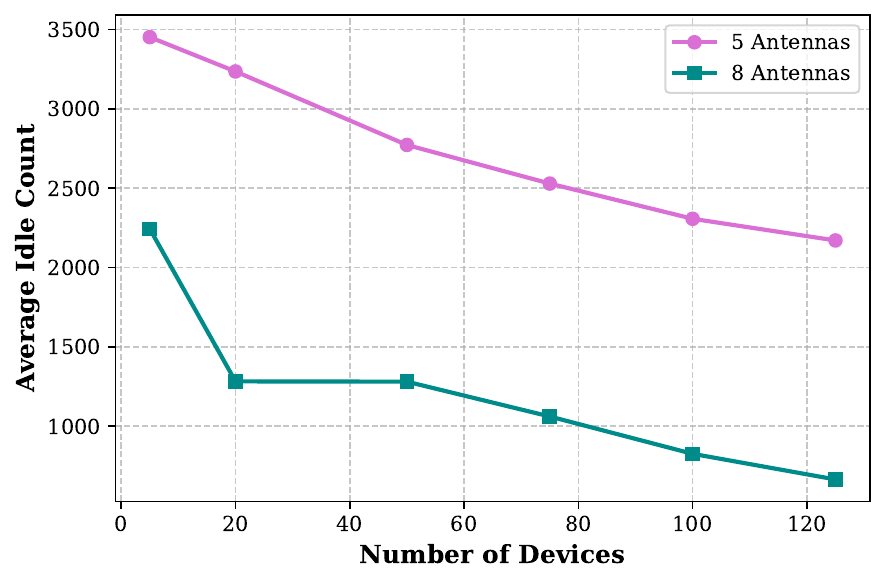}
    \caption{Average number of idle slots (averaged across all device placements for last 100 episodes) versus network size for $M = 5$ and $M = 8$ antennas using the ADRQN policy with $H = 50$.}
    \label{fig:idle}
\end{figure}

\begin{figure}[t]
    \centering
    \includegraphics[width=\columnwidth]{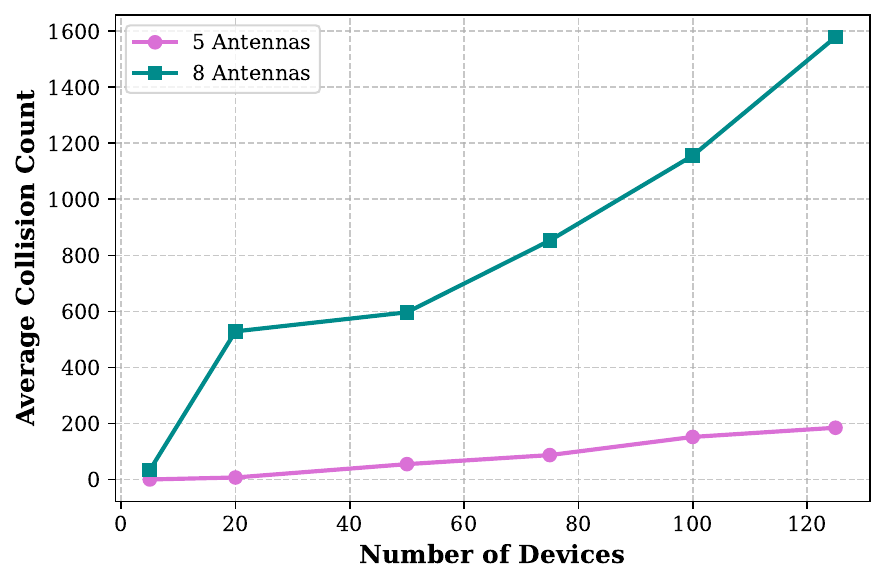}
    \caption{Average number of collision slots (averaged across all device placements for last 100 episodes) versus network size for $M = 5$ and $M = 8$ antennas using the ADRQN policy with $H = 50$.}
    \label{fig:collision}
\end{figure}

Figs~\ref{fig:idle} and~\ref{fig:collision} show the average idle and collision slot counts for the  ET/APs with $M = 5$ and $M = 8$ antennas, respectively, with the number of devices in the network is increased from $N = 5$ to $N = 125$ BEHDs. The simulations were run with the proposed ADRQN based policy with history length $H = 50$. The idle slots monotonically decrease as the number of devices in the network is increased, whereas the number of collisions is increased for both antenna configurations. The 5-antenna ET/AP configuration maintains higher idle slot counts as the number of devices is increased from $N = 5$ to $N = 125$, ranging from 3451 to 2170, whereas for 8-antenna configuration the decrease is from 2243 to 664 over the same range. The 8-antenna configuration also maintains significantly higher numbers of collisions (e.g. 852 at $N = 75$, compared to the 5-antenna configuration which is 87 only). This is due to the fact that the 8-antenna configuration maintains narrower beamwidths and thus focuses the transmitted power in smaller angular ranges. This enables multiple spatially proximate devices \emph{in a dense network} to charge rapidly and reach the transmission threshold $Q_{\mathrm{th}}$ within the same or adjacent slots, thereby increasing the likelihood of simultaneous transmissions and resulting collisions. On the other hand, the 5-antenna configuration maintains wider beamwidths that result in the dissipation of transmitted power over larger angular ranges. This in turn staggers the time it takes for devices of nearby devices to reach threshold $Q_{\mathrm{th}}$, reducing collisions at the cost of increased idle time.

\begin{figure}[t]
    \centering
    \includegraphics[width=\columnwidth]{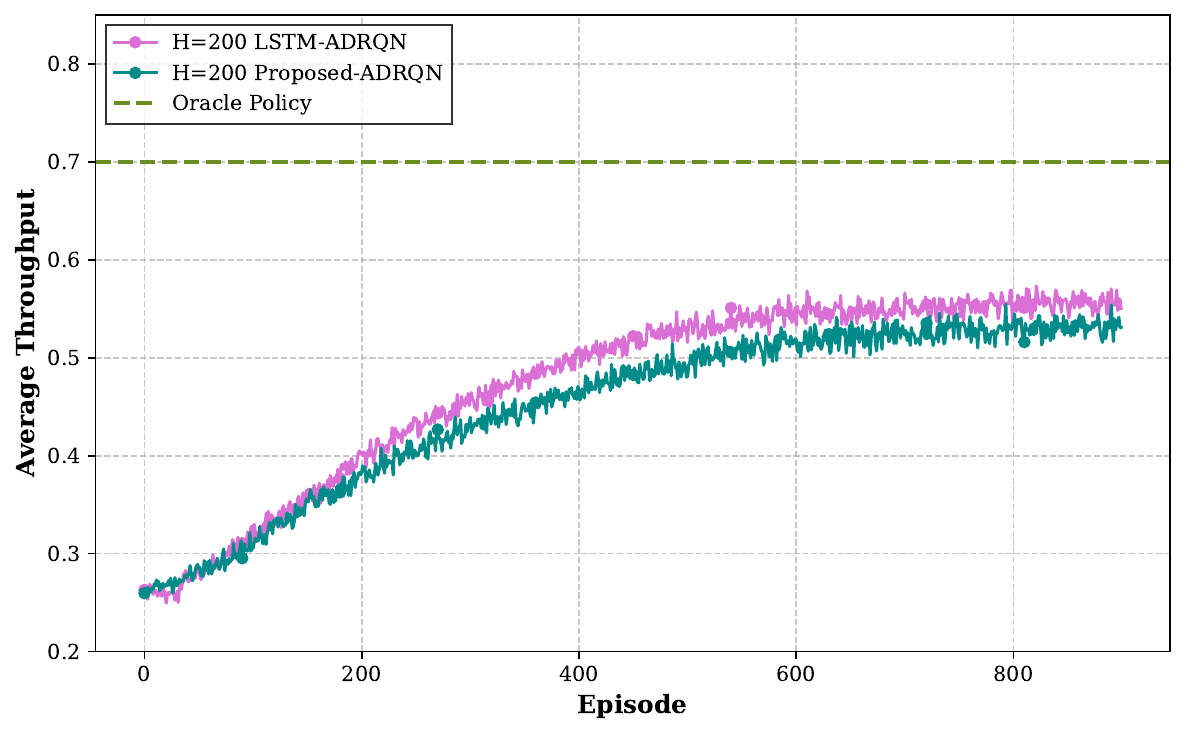}
    \caption{Convergence of the proposed ADRQN and LSTM-ADRQN ($H = 200$, $N = 50$, $\xi=0.85$, $M = 8$) over 900 training episodes. The dashed line denotes the oracle policy throughput with full state knowledge.}
    \label{fig:convergence_oracle}
\end{figure}

As presented in Fig.~\ref{fig:convergence_oracle}, we plot the throughput with respect to the episodes for our proposed ADRQN (RNN) and its LSTM version, with history length $H = 200$, for a system consisting of $N = 50$ BEHDs and $M = 8$ antennas. The horizontal dashed line in the plot represents the throughput value achieved by the oracle policy described in Section ~\ref{sec:oracle}. The learning curves for both recurrent neural networks show that average system throughput improves steadily over time. The initial average throughput is around 0.25, and it improves to values greater than 0.52 for both ADRQN (RNN) and LSTM, as training progresses through 900 episodes. This monotonic improvement in average system throughput over time indicates that neither ADRQN (RNN) nor its LSTM version overfits to early exploration patterns and continues to refine its beam-steering policies. Also, it is important to observe that both learning curves remain below the oracle value, which is around 0.70. This is expected, as our proposed oracle policy utilizes perfect knowledge of charge states, channel realizations, and system states, which is not available to our learning-based policies. The gap between the converged ADRQN throughput and oracle value represents the cost of partial observability in our problem formulation. Our learning-based policies achieve 75-80\% of oracle performance without channel estimation, charge-level reporting, and device state tracking, and only relying on macro-level slot feedback.

\begin{figure}[t]
    \centering
    \includegraphics[width=\columnwidth]{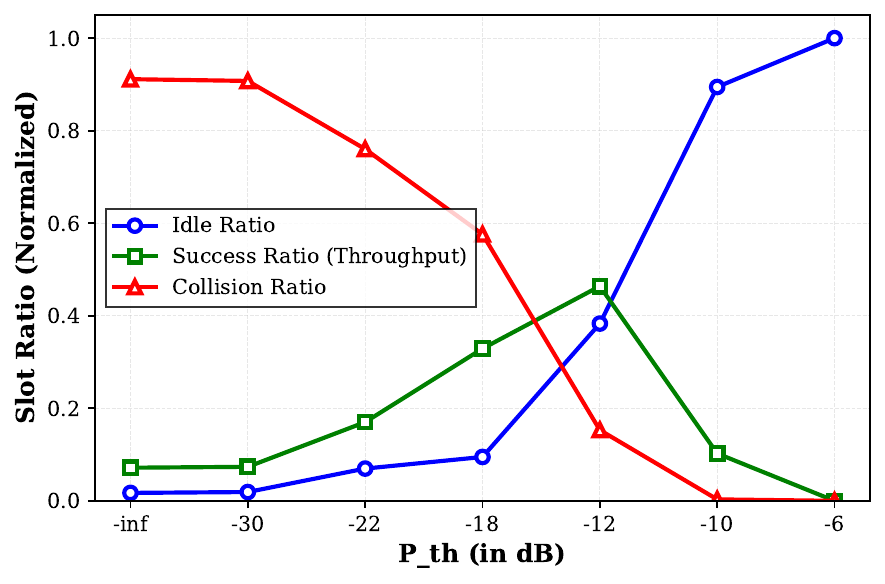}
    \caption{Normalised idle, success, and collision slot ratios versus the power threshold $P_{\mathrm{th}}$ (in dB relative to $P_T$), for $N = 50$ and $M = 8$. The success ratio (throughput) peaks at $P_{\mathrm{th}} = -12$~dB.}
    \label{fig:pth_sweep}
\end{figure}

Fig.~\ref{fig:pth_sweep} shows the normalised slot ratios (idle, success, and collision) as a function of the eligibility threshold $P_{\mathrm{th}}$ varied over $\{-6, -10, -12, -18, -22, -30, -\infty\}$~dB (relative to $P_T$), for $N = 50$ EH devices, $M = 8$ antennas and $H=50$ history sequence length. The success ratio corresponds to the average throughput, while idle and collision ratios are normalised by the total number of slots (e.g. idle ratio = total number of idle slots/total number of slots ). 

\subsection{Discussion}\label{subsec:discussion}

The experimental results show that a larger history size results in better performance. This confirms the main argument that, even though the ET/AP cannot directly observe the charge states of the devices, it can infer this information from previous interactions. Of the three architectures, the feedforward DQN uses a fixed-dimensional vector, where the size of the vector grows linearly with $H$ but eventually exceeds the network's capacity. The RNN-based ADRQN architecture uses long-range dependencies in a fixed-dimensional vector. The performance of this architecture is improved in the LSTM architecture, showing its capability for better inference of the long charging cycles in the WPCN scenario. Furthermore, tabular Q-learning is infeasible for this problem because the observation-history space grows exponentially as $(S \times |\mathcal{O}|)^H$, exceeding $10^{144}$ distinct states for the configurations considered in this work, necessitating function approximation.

The observed throughput scaling with the size of the network is in line with the expected performance of slotted ALOHA. As the number of devices increases, the slot utilization improves initially; however, beyond a certain point, it is dominated by collisions. 
The performance difference between the ADRQN and the oracle represents the price of partial observability. The optimal oracle could be computed using dynamic programming (DP), but the $N$-dimensional charge state space is too large to be precisely computed (for example, discretising each device's charge level into 10 bins between $Q_0$ and $Q_{\mathrm{th}}$ yields $10^{50}$ states for $N = 50$ devices), making the exact DP intractable. The oracle proposed in this work, which has a per-slot cost of $O(S^k \times k \times N)$, offers a robust upper bound. The proposed ADRQN-based policy performs fairly well compared to the oracle without global network information. 

It is also noteworthy that the proposed policy does not explicitly guarantee fairness among devices. However, as illustrated in Figs.~\ref{fig:coverage_5} and~\ref{fig:coverage_8}, the beam directions are chosen such that their coverage regions span nearly the entire $360^\circ$ angular range, ensuring almost every device falls within the main lobe of at least one beam. Devices located at the edges of these coverage zones may receive relatively less power, but they do not face complete energy starvation, as they continue to receive energy through \text{NLoS} channel components and side-lobe radiation even when the beam is steered in a different direction.

\section{Conclusion}\label{sec:conclusion}

Our work introduced a new design paradigm for MAC protocols in networks comprising battery-free energy harvesting devices. We proposed an approach to maximize network throughput by intelligently steering the energy beam from an energy transmitter-cum-access point. To preserve scarce harvested energy, our approach eliminated the overheads inherent in conventional MAC protocols. By formulating the joint problem of beam steering and random access under slotted ALOHA as a Partially Observable Markov Decision Process (POMDP), we developed an Action-specific Deep Recurrent Q-Network (ADRQN) framework that leveraged the history of ternary slot outcomes and beam directions to dynamically determine the optimal next beam direction.

We benchmarked our proposed approach and its variants against traditional non-learning policies and an oracle policy with access to the global knowledge of the network. The numerical results clearly showed that not only did the proposed approach improve throughput by up to 68\% compared to round-robin and random selection policies, but it also achieved up to 80\% of the throughput of the oracle policy despite having access to limited system information. 


Future research opportunities arise from generalizing the considered system model, specifically through the inclusion of heterogeneous energy harvesting devices or the implementation of a more complex energy transfer infrastructure. The heterogeneity in energy harvesting devices could be attributed to differences in structure (e.g., storage capacity, number of antennas), operation (e.g., backscattering capability), or both. On the other hand, the energy transfer infrastructure could comprise multiple energy transmitters, each capable of transmitting multicarrier energy signals and dynamically regulating transmit power as the beam direction changes. Finally, incorporating optimization goals that prioritize fairness and maintain strict limits on the age-of-information (AoI) could help address a broader spectrum of application requirements.


\ifCLASSOPTIONcaptionsoff
  \newpage
\fi

\bibliographystyle{IEEEtran}
\bibliography{References/EHMAC}

\end{document}